# Disentangling structural and kinetic components of the α-relaxation in supercooled metallic liquids

*Nico Neuber*[a*], Oliver Gross[a,b], Maximilian Frey[a], Benedikt Bochtler[a,b], Alexander Kuball[a,b], Simon Hechler[a,b], Fan Yang[c], Eloi Pineda[d], Fabian Westermeier[e], Michael Sprung[e], Isabella Gallino[a], Ralf Busch[a] and Beatrice Ruta[f*]

[a]*Chair of Metallic Materials, Saarland University, Campus C6.3, 66123 Saarbrücken, Germany*

[b]*Amorphous Metal Solutions GmbH, Michellinstraße 9, 66424 Homburg, Germany*

[c]*Institut für Materialphysik im Weltraum, Deutsches Zentrum für Luft- und Raumfahrt (DLR), 51170 Köln, Germany*

[d]*Department of Physics, Institute of Energy Technologies, Universitat Politècnica de Catalunya - BarcelonaTech, 08019 Barcelona, Spain*

[e] *Deutsches Elektronen-Synchrotron DESY, Notkestr. 85, 22607 Hamburg, Germany*

[f] *Université Lyon, Université Claude Bernard Lyon 1, CNRS, Institut Lumière Matière, Campus LyonTech - La Doua, F-69622, Lyon, France*

*corresponding authors*

## Abstract

The particle motion associated to the α-relaxation in supercooled liquids is still challenging scientists due to its difficulty to be probed experimentally. By combining synchrotron techniques, we found the existence of microscopic structure-dynamics relationships in $Pt_{42.5}Cu_{27}Ni_{9.5}P_{21}$ and $Pd_{42.5}Cu_{27}Ni_{9.5}P_{21}$ liquids which allows us to disentangle structural and kinetic contributions to the α-process. While the two alloys show similar kinetic fragilities, their structural fragilities differ and correlate with the temperature dependence of the stretching parameter describing the decay of the density fluctuations. This implies that the evolution of dynamical heterogeneities in supercooled alloys is determined by the rigidity of the melt structure. We find also that the atomic motion not only reflects the topological order but also the chemical short-range order, which can lead to a surprising slowdown of the α-process at the mesoscopic length scale. These results will contribute to the comprehension of the glass transition, which is still missing.

# Introduction

Despite decades of studies, glass formers keep fascinating scientists and are often considered as archetypes of complex systems[1–5]. Understanding glass formation and the microscopic mechanisms responsible for the extreme slowing down of the structural α-relaxation process in supercooled liquids, prior to the glass transition, is a challenging task due to the difficulty associated with the measurement of the atomic motion using experiments and numerical simulations. Despite these challenges, several common features have been identified in recent years: i) the glass transition is often considered as a dynamical process accompanied by weak structural changes [6], although indications for underlying increasing structural correlation lengths have been reported in the last ten years in numerical studies [7,8]; ii) during undercooling, the relaxation time of the α-process (or the viscosity) of glass formers increases over several orders of magnitude in a faster-than-exponential way, until the liquid eventually transforms in an out-of-equilibrium glass [9]; iii) the supercooled liquid is characterized by the emergence of dynamical heterogeneities, which result in a stretched exponential long time decay of the density fluctuations associated to the α-relaxation [3,10].

Usually, the evolution of the dynamics in supercooled liquids approaching the glass transition temperature, $T_g$, is described by the kinetic fragility, m, a parameter introduced for the first time by C. A. Angell[11]. The kinetic fragility quantifies the departure from an Arrhenius behavior of the shear viscosity – or the structural relaxation time, τ, upon cooling a liquid at $T_g$. Following this description, all those liquids exhibiting an Arrhenius-like evolution of the viscosity on cooling, often characterized by strong directional bonds, are called strong, while fragile liquids display a super Arrhenius behavior and are usually systems with non-directional bonds or van der Waal interactions. In the last years, many correlations have been proposed between the kinetic fragility and several properties of both liquids and glasses, highlighting the importance of this parameter[12–14].

By combining state-of-the-art synchrotron techniques, X-ray Photon Correlation Spectroscopy (XPCS) and High Energy X-ray diffraction (HEXRD), we show here that the evolution of the atomic motion, and therefore of the structural α-relaxation process, on approaching the glass transition should be described not only by the kinetic fragility but also by the structural fragility of the liquid[15]. The latter parameter is related to the evolution of the structure on the length scale of medium range order, evolving similarly to the degree of dynamical heterogeneities emerging during the cooling. The results of this work emphasize the importance of the subtle changes accompanying the vitrification process by showing that both the temperature and length scale dependence of the microscopic α-relaxation are correlated with the underlying structural changes.

Metallic liquids are ideal candidates for this study. Due to their simple metallic bond structure, they have only translational motion and, therefore, they can be considered as model systems to study the glass transition. The two metallic glass-forming liquids $Pt_{42.5}Cu_{27}Ni_{9.5}P_{21}$ and $Pd_{42.5}Cu_{27}Ni_{9.5}P_{21}$ were selected for this study for several more reasons. According to structural models of glass-forming liquids, based on the efficient packing of representative structural units (clusters) [16], elements with atomic radii that deviate no more than 2 % are considered as topologically equivalent [17], which is true for Pt and Pd in these two compositions. However, HEXRD experiments have revealed significant differences in their structures [18]. Based on these experiments, it is suggested that the structure of both liquids is dominated by two different structural motifs. While the $Pd_{42.5}Cu_{27}Ni_{9.5}P_{21}$ liquid consists mainly of icosahedral motifs, leading to a pronounced short-range order (SRO), the structure of the $Pt_{42.5}Cu_{27}Ni_{9.5}P_{21}$ shows a larger part of trigonal prisms, leading to a rather pronounced medium range order (MRO). This MRO is manifested in the appearance of a pre-peak in the low-Q range of the total structure factor S(Q), before the first sharp diffraction peak (FSDP), which is present in the Pt-based liquid but absent in the Pd-based one. In contrast, the Pd-based liquid shows a shoulder on the larger Q-side of the second sharp diffraction peak, associated to icosahedral SRO (see SI Fig. S1). Despite these structural differences, both alloys show similar kinetic fragilities [19–21], whereas the excess heat capacity around the glass transition, as well as melting enthalpy and entropy are found to be much higher in the Pt-based liquid [21–23] than in the Pd-based alloy. This indicates that the Pt-based alloy is more fragile from the thermodynamic point of view. Furthermore, Pt-based glasses tend to show a more ductile mechanical performance than their Pd-based peers, which, in contrast, are more sensitive to cooling rate and annealing related embrittlement [24–26]. Finally, both systems have an excellent glass-forming ability (GFA), as described by the critical casting thickness $d_c$ which is $d_c = 20$ mm for $Pt_{42.5}Cu_{27}Ni_{9.5}P_{21}$ and $d_c = 80$ mm for the $Pd_{42.5}Cu_{30}Ni_{7.5}P_{20}$ [27,28]. Such exceptional resistance to crystallization allows us to perform dynamical measurements in the supercooled liquid phase. All these features make these two systems perfect candidates to study the connection between atomic motion and the inherent structures.

## Results and Discussion

*Collective particle motion in the supercooled liquid phase*

Information on the dynamics can be obtained from the intermediate scattering function (ISF), $f(Q,t)$, which describes the temporal decay of the particle density fluctuations. This quantity can be obtained in an XPCS measurement by the determination of the correlation function of the intensity fluctuations, $g_2(Q,t)$, generated by the scattering of coherent X-rays from the sample, being $g_2(Q,t) = 1+\gamma \cdot |f(Q,t)|^2$. In this expression, $\gamma$ is the experimental contrast and $Q$ the probed wave-vector [29]. In glass-formers, $g_2(Q,t)$ can be described by the Kohlrausch-Williams-Watts (KWW) function $g_2(Q,t) = 1+c \cdot \exp[-2(t/\tau)^\beta]$ where $c = \gamma^* f_q^2$ is the product between the experimental contrast and the square of the nonergodicity parameter, $f_q$, of the system, $\tau$ is the relaxation time, and $\beta$ the shape parameter [29]. By definition, the shape parameter describes the degree of nonexponentiality of the ISF, which originates from the underlying distribution of relaxation times, and it can therefore be considered as a measure of the heterogeneous nature of the dynamics in the liquid phase.

Fig. 1a) and b) show the temperature dependence of normalized $[g_2(Q,t)-1]/c$ functions measured at $Q_{FSDP} \approx 2.8$ Å$^{-1}$, corresponding to the position of the first sharp diffraction peak (FSDP) in the static profile. For both compositions, the data have been acquired in the supercooled liquid phase, cooling from about $(T_g+25)$ K, with $T_g$ (Pt$_{42.5}$Cu$_{27}$Ni$_{9.5}$P$_{21}$) = 506 K and $T_g$ (Pd$_{42.5}$Cu$_{27}$Ni$_{9.5}$P$_{21}$) = 566 K for the applied cooling rate of 0.025 K s$^{-1}$. By decreasing the temperature by only $\approx$ 20 K, the decay of the curves clearly shifts by about two orders of magnitude towards longer times, indicating the rapid slow-down of α-relaxation process when approaching the glass transition. The corresponding relaxation times display a very similar evolution with temperature in both compositions. This is shown in Fig. 1c) and d), where we report the mean relaxation time $\langle \tau(Q) \rangle = \Gamma(\beta(Q)^{-1}) \frac{\tau(Q)}{\beta(Q)}$, where $\Gamma$ is the Gamma function [30]. In both systems, $\langle \tau(Q,T) \rangle$ can be described by the Vogel-Fulcher-Tammann (VFT) equation with $\langle \tau(Q,T) \rangle = \tau_0 \cdot \exp[(D^*T_0)/(T-T_0)]$ using the parameters obtained by macroscopic studies such as viscosity measurements, from Ref. [19] for the Pt-alloy and from new measurements (Fig. S2 in the SI) for the Pd-alloy.

In contrast to $\langle \tau(Q,T) \rangle$, the shape of the ISFs exhibits a different evolution with temperature in the two liquids. In both cases, the curves can be described by a stretched exponential decay with a KWW parameter $\beta(Q,T) < 1$, indicating the heterogeneous nature of the dynamics in the viscous liquid phase (Fig. 1e) and f)) [3]. However, for the Pd-liquid $\beta(Q_{FSDP},T)$ remains constant at a value of $\approx 0.55$ within the probed temperature range, whereas for the Pt-liquid,

$\beta(Q_{FSDP},T) \approx 0.7$ at high temperature and then clearly decreases by about 20% of its initial value during cooling. Similar temperature dependences of the KWW parameters have been observed also at length scales corresponding to few interatomic distances (see Fig. S3 in SI). The rapid decrease of $\beta(Q,T)$ on cooling in the Pt-based alloy implies the failure of the time-temperature superposition principle for this composition (see also Fig. S4 in SI). This result contrasts with previous XPCS studies of metallic glass formers where $\beta(T)$ has been found T-independent in all measured supercooled liquids allowing to superpose all the correlation curves in a single master curve [31–33]. Although the relationship between macroscopic and microscopic dynamics is not straightforward, also most metallic glasses show a T-independent shape of the α-relaxation peak in mechanical experiments. The shape of such peak is in agreement with a KWW exponent of around 0.5[34–36]. This means that the satisfaction of the time-temperature-superposition (TTS) principle is expected in many metallic glass-forming systems. Further detailed studies on the dynamic relaxation processes in metallic glasses can be found in Refs. [37–41].

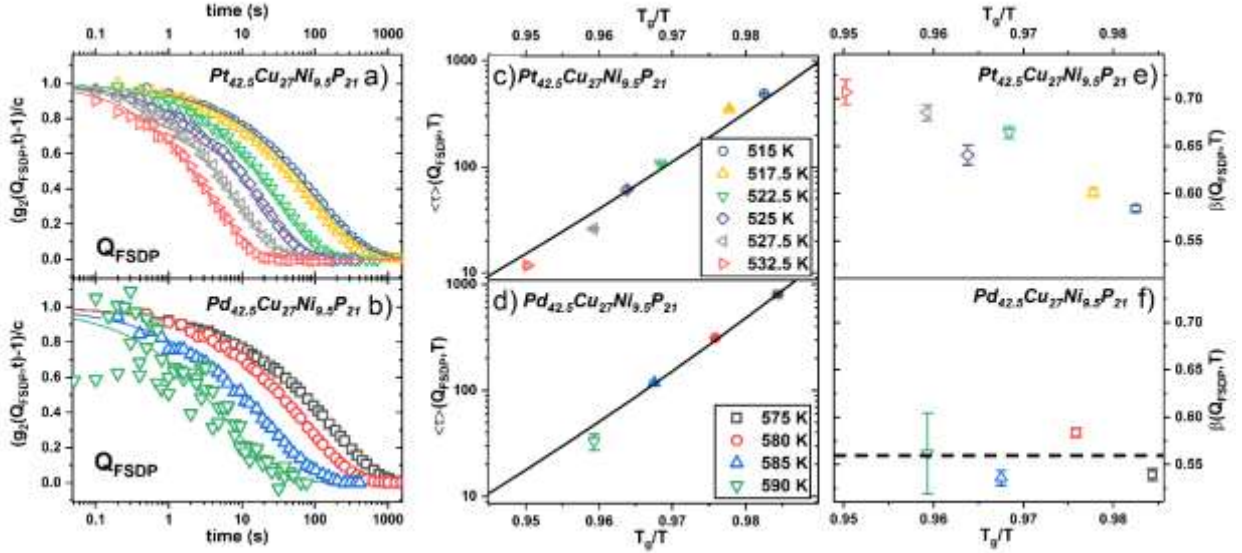

**Figure 1: Temperature dependence of the atomic dynamics in $Pt_{42.5}Cu_{27}Ni_{9.5}P_{21}$ and $Pd_{42.5}Cu_{27}Ni_{9.5}P_{21}$ supercooled liquids.** **(a-b)** Normalized intensity autocorrelation functions measured at various temperatures for $Pt_{42.5}Cu_{27}Ni_{9.5}P_{21}$ **(a)**, and $Pd_{42.5}Cu_{27}Ni_{9.5}P_{21}$ **(b)** at the position of the FSDP ($Q_{FSDP} \approx 2.8$ Å$^{-1}$ for both systems). **(c-d)** $<\tau(Q,T)>$ of the Pt- and Pd-based alloys measured at $Q_{FSDP}$. The solid lines are fits to the VFT equation (see text). **(e-f)** Temperature dependence of the corresponding $\beta(Q,T)$ parameter for $Pt_{42.5}Cu_{27}Ni_{9.5}P_{21}$ **(e)** and $Pd_{42.5}Cu_{27}Ni_{9.5}P_{21}$ **(f)**. The dashed line is a guide to the eyes.

*Structural evolution in the supercooled liquid phase*

The different temperature evolution of the dynamics in the two alloys can be explained by looking at the temperature dependence of the underlying structure. Figure 2 shows the temperature dependence of the peak intensity (a), full width at half maximum (FWHM) (b) and peak position (c) of the FSDP of the static structure factors S(Q) measured in both compositions with HEXRD (the corresponding diffraction spectra can be found in SI Fig. S1, S5 and S6). All data are linearly fitted in the supercooled liquid state and rescaled to the $T_g$ at the measured rate of 0.33 K s$^{-1}$. While the peak intensity displays a similar temperature dependence in both alloys (Fig. 2a)), the FWHM and the peak position exhibit a steeper evolution with temperature in supercooled $Pt_{42.5}Cu_{27}Ni_{9.5}P_{21}$ (b) and c)). Structural studies have shown that the FWHM of the FSDP correlates well with the correlation length over which the period of a repeated unit persists, underlying its connection to the medium range order in non-crystalline systems [42]. The more pronounced temperature dependence of FWHM$_{FSDP}$ of the Pt-liquid (Fig. 2b)) thus provides evidence of the presence of a more pronounced reorganization of the MRO in the deeply undercooled Pt-liquid compared to the Pd-based one. This effect is accompanied by a more rapid evolution of the peak position with temperature (Fig. 2c)) in the $Pt_{42.5}Cu_{27}Ni_{9.5}P_{21}$ supercooled liquid, confirming the presence of significant structural rearrangements during cooling.

The larger tendency to temperature induced structural rearrangements at the MRO length scale in the $Pt_{42.5}Cu_{27}Ni_{9.5}P_{21}$ liquid is likely responsible for the temperature evolution of the KWW exponent describing the decay of the ISFs. At temperatures (much) higher than those probed in this work, the system is likely governed by diffusion with ISFs described by a single exponential decay (i.e., β = 1) and τ~Q$^{-2}$. During cooling in the supercooled liquid phase, the viscosity increases (Fig. S2), the dynamics slows down, and β(T) decreases due to the formation of cages of atoms and the occurrence of dynamical heterogeneities. The relatively large value of β ≈ 0.7 in the Pt-alloy at only 25 K above $T_g$, suggests that the cages can already break easier at this temperature leading to more homogeneous dynamics. During cooling, the liquid experiences a rapid slowdown of the dynamics which hinders the particle motion and cages break less frequently, leading to an increased heterogeneity in the dynamics and thus to a lower value of β in the Pt-alloy. The low constant value of the shape parameter in the $Pd_{42.5}Cu_{27}Ni_{9.5}P_{21}$ and the absence of important structural reorganizations suggest instead that this supercooled liquid is already in a dynamically highly heterogeneous, stable state. In the Pd-liquid, the transition from diffusive dynamics by collective motion at high temperature to activated collective dynamics in the supercooled liquid phase, if present, probably occurs at temperatures much higher than those measured in the current experiment. This is in agreement with neutron studies in other Pd-based alloys which report nonexponential decays of the ISFs even in the high temperature melt [43].

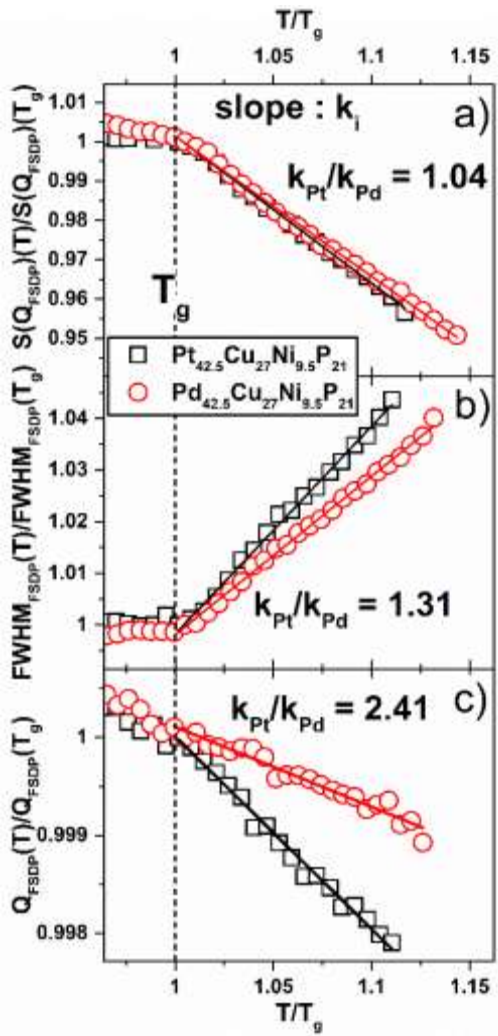

**Figure 2: Temperature dependence of the FSDP of $Pt_{42.5}Cu_{27}Ni_{9.5}P_{21}$ and $Pd_{42.5}Cu_{27}Ni_{9.5}P_{21}$ measured with synchrotron XRD.** Peak intensity $S(Q_{FSDP})$ **(a)**, full width at half maximum $FWHM_{FSDP}$ **(b)** and peak position ($Q_{FSDP}$) **(c)** of the FSDP as a function of $T/T_g$ for the Pt-based (black squares) and Pd-based (red circles) alloys. All data are normalized to the respective value of the glass transition. In all panels, k is the respective slope of a linear fit of the data in the liquid state of the Pt/Pd-based alloys (above $T_g$).

*Kinetic and structural fragility*

For a better quantitative comparison of the relaxation behavior of both liquids, Fig. 3a) shows a fragility plot where we report the inverse temperature as a function of the average relaxation time $<\tau(Q_{FSDP})>$. On the atomic level, both systems exhibit similar kinetic fragilities, *m*, with $m_{Pt} = 53.1 \pm 2.6$ and $m_{Pd} = 55.3 \pm 3.2$. Here, *m* is defined as the logarithms slope of the kinetic variable X (e.g. viscosity/relaxation time) at $T_g$ via $m = d\log X/d(T_g/T)|_{T=T_g}$ [9,11]. These values are in good agreement with those obtained from macroscopic measurements [19,44,45], using the relation $m = 16 + 590/D^*$, which gives $m_{Pt} = 54.5$ ($D^* = 15.3$) and $m_{Pd} = 57$ ($D^* = 14.5$) (compare SI Table 1).

Despite their similar kinetic fragility, the evolution of the α-relaxation process during cooling is not the same in the two systems, as seen by the different values and temperature dependencies of the KWW shape parameter. This difference seems to disappear on approaching $T_g$ where both liquids reach a similar degree of nonexponentiality of the ISF (Fig. 3b)). A similar behavior was observed in macroscopic measurements of ultraphosphate liquids with different sodium content, which display different temperature evolutions of β(T) merging all to a value of β($T_g$) ≈ 0.5 [46]. We find that the rate of change of β($Q_{FSDP}$,T) with <τ($Q_{FSDP}$,T)>, i.e., the slope dβ/dlog(<τ>), is about -0.08 for the Pt-based alloy. This value is comparable with the steepest slope observed for pure ultraphosphate (dβ/dlog(<τ>) ~ -0.06, [46]) where it has been attributed to a high degree of cooperativity in the 3D-network of $PO_4$ tetrahedra [46]. In our case, the steep evolution of β(T) appears to be related to the tendency of the Pt-alloy to temperature induced structural rearrangements on the length-scale of MRO, which itself is connected to fragility of metallic systems [15] (see Fig. 2). In amorphous metallic systems, the MRO is in fact mainly associated to the FSDP which is representative of structural rearrangements occurring on length scales beyond r > 6 Å [47]. The evolution of the 3$^{rd}$ ($r_3$) and 4$^{th}$ peak ($r_4$) of the reduced pair-distribution function G(r) provides therefore another tool to gain insights into the structural changes of the two liquids [15]. Fig. 3c) shows the temperature dependence of the relative change of the volume expansion $\Delta V_{4-3} = 4/3\pi(r_4^3-r_3^3)$ of a shell between $r_3$ and $r_4$ with respect to the value at $T_g$. As it can be immediately deduced from Fig. 3c), the ratio $\Delta V_{4-3}/\Delta V_{4-3}(T_g)$ mirrors the behavior of β($Q_{FSDP}$,T) suggesting a possible correlation between changes on the MRO length scale and the temperature evolution of the KWW parameter. Following Ref. [15], we can define a structural fragility, $m_{str4-3}$, from the slope of $\Delta V_{4-3}/\Delta V_{4-3}(T_g)$ and we find $m_{str4-3,Pt} = 0.035 \pm 0.004$ for the $Pt_{42.5}Cu_{27}Ni_{9.5}P_{21}$ and a basically temperature independent $m_{str4-3,Pd} \approx 0.005 \pm 0.006$ for the $Pd_{42.5}Cu_{27}Ni_{9.5}P_{21}$. Both values are in good agreement with those given in Refs. [45] and [15]. In Ref. [15] $m_{str4-3}$ was found to correlate to the kinetic fragility of the liquid through the empirical expression $m = ((m_{str4-3}+0.124)/2.95\times10^{-3})$. While this relation is fulfilled for the Pt-liquid being $m_{Pt} = 53.9 \pm 1.3$ as well as numerous Zr-based liquids, it fails for the Pd-based alloy as it gives $m_{Pd} = 42.6 \pm 3.3$, which is significantly lower than the measured value. The failure of the proposed correlation for the Pd-based alloy, the temperature independent KWW stretching exponent (Fig. 3b)) and the weak temperature evolution on the

MRO length-scale (Fig. 2), support the idea of a smaller tendency to structural rearrangements and thus of a rather strong structural behavior of the Pd-based alloy at the probed length scales, in agreement with the estimation obtained from the structural fragility. At this point it should be noted that further support for the structural strong behavior of the Pd-based liquid is provided by the behavior of the specific isobaric heat capacity $c_p$. During undercooling of the Pt-liquid, a more rapid rise of $c_p$ can be observed compared to the Pd-based liquid [22,23]. This change of the liquid $c_p$ is connected to the rate of the loss in the excess entropy, which itself is proportional to the configurational entropy of the liquid [13,21,48]. Within the framework of Adam and Gibbs [49], this behavior can be seen as the thermal signature of a faster growth of cooperatively rearranging regions in the Pt-liquid with ongoing undercooling. Thus, it further confirms the larger tendency of the Pt-liquid to structural changes indicated by the steep temperature dependence of the FSDP (Fig. 2), the evolution of the degree of dynamical heterogeneity $\beta(Q_{FSDP},T)$ (Fig. 3b)), and the fragile value of the structural fragility (Fig. 3c)).

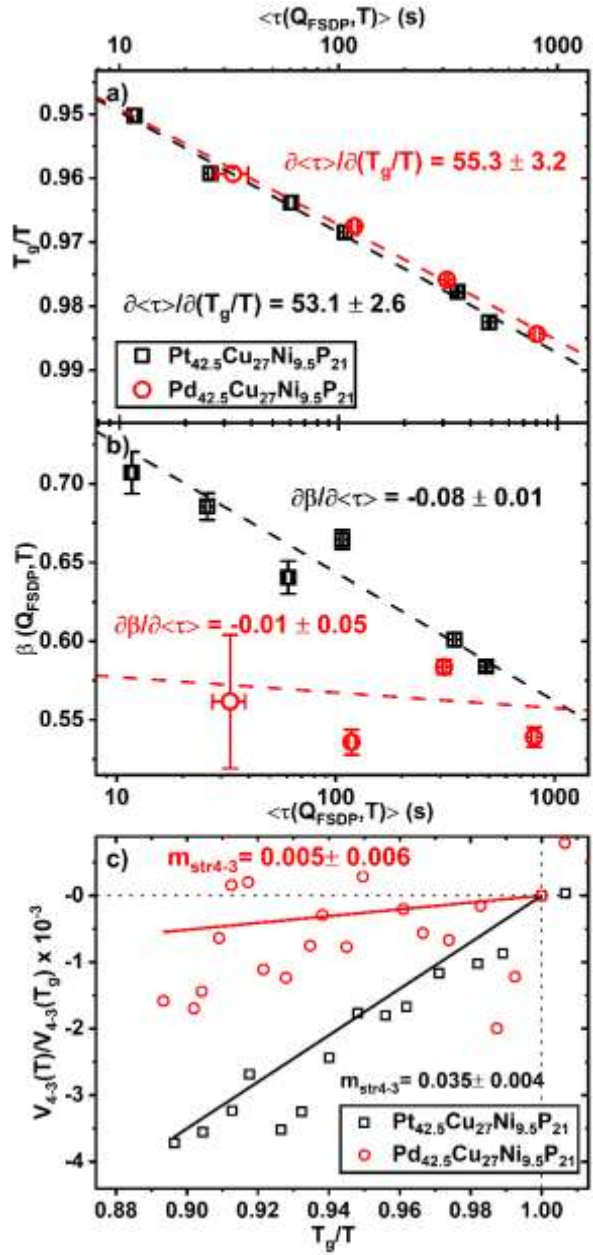

**Figure 3: Temperature evolution of the dynamical parameters and of the structure on the MRO length scale. (a)** $T_g$-scaled (for a cooling rate of 1.5 K min$^{-1}$) fragility plot of the mean relaxation time $<\tau(Q_{FSDP})>$ for Pt$_{42.5}$Cu$_{27}$Ni$_{9.5}$P$_{21}$ and Pd$_{42.5}$Cu$_{27}$Ni$_{9.5}$P$_{21}$ (black squares and red circles, respectively). Solid lines are linear fits on the logarithmic scale underlining the similar kinetic fragility of the two liquids. Note that the slope provided in the figure comes from the conventional definition of the kinetic fragility, *m*, where the axes are inverted. **(b)** Corresponding $\beta(Q_{FSDP},T)$ as a function of $<\tau(Q_{FSDP})>$. The dashed lines are linear fits of the data on the logarithmic scale. **(c)** Evolution of $V_{4-3}(T)/V_{4-3}(T_g)$ as a function of $T/T_g$, where $T_g$ is the glass transition temperature during heating with a rate of 0.33 K s$^{-1}$. A linear fit (full line in respective color) of the data is used to calculate the structural fragility $m_{str4-3}$ of the liquid phase.

*Wave-vector dependence of the dynamics*

The different atomic dynamics of the two alloys is even more evident by looking at the wave-vector dependence of the relaxation time. As shown in Fig. 4a), $\tau(Q)$ displays two maxima for the $Pt_{42.5}Cu_{27}Ni_{9.5}P_{21}$ liquid at $Q_{PP} = 2.05$ Å$^{-1}$ and $Q_{FSDP} = 2.8$ Å$^{-1}$. These values correspond to the position of the maximum of a structural pre-peak ($Q_{PP}$) and of the main FSDP observed in the corresponding static structure factor (dashed grey line). In contrast, $\tau(Q)$ exhibits only one maximum at $Q_{FSDP}$ for $Pd_{42.5}Cu_{27}Ni_{9.5}P_{21}$, agreeing with the corresponding $S(Q)$ (Fig. 4b)). The increase of $\tau(Q)$ in correspondence with the different underlying structural motifs implies the existence of more stable configurations at the corresponding length scales. While the maximum at $Q_{FSDP}$ is typical of glass formers, usually observed in high temperature liquids by neutron scattering (de Gennes narrowing [50,51]), the prominent increase of $\tau(Q)$ at $Q_{PP}$ in the $Pt_{42.5}Cu_{27}Ni_{9.5}P_{21}$ is more surprising due to the very weak intensity of the pre-peak in the static profile.

A similar evolution with Q is observed also for $\beta(Q)$ in both supercooled liquids (Fig. 4c) and d)). For $Pt_{42.5}Cu_{27}Ni_{9.5}P_{21}$, however, the maximum at the pre-peak is broader and less pronounced compared to that in $\tau(Q)$ (Fig. 4c)). The larger values of $\beta(Q)$ in the Pt-alloy at the positions of the of structural maxima suggest more homogeneous dynamics at these length scales with respect to neighboring Q values. To better understand the observed behavior, we can distinguish two zones: the atomic scale, i.e., in the proximity of the FSDP for 2.5 Å$^{-1}$ < Q < 3.2 Å$^{-1}$; and the intermediate regime at a length scale of few interatomic distances for 1.5 Å$^{-1}$ < Q < 2.5 Å$^{-1}$.

We first concentrate on the atomic scale. Besides the experimental results reported for liquids at high temperature [50,51], a similar evolution of the collective relaxation time with Q has been observed also in numerical simulations of different glass formers close to the glass transition like, for instance, supercooled silica, water and hard spheres [52–54] where it has been described within the mode coupling theory. It should be noted that while in network glass formers $\beta(Q)$ is found constant with Q, only in hard spheres $\beta(Q)$ correlates also with the intensity evolution of the $S(Q)$ and shows a maximum at the FSDP [53,55] as in our data. This suggests a stronger correlation between metallic glasses and hard spheres systems [56].

In the intermediate regime, the presence of a second peak in the relaxation time corresponds to an additional slow-down of the collective motion in correspondence with the weak structural pre-peak at a few interparticle distances. Similar behaviors on the MRO length scale have been reported only in few other glass formers as silicates, ortho-terphenyl, polymeric systems and $[Ca(NO_3)_2]_{0.4}[KNO_3]_{0.6}$ (CKN) [57–60], where it has been associated to the underlying topological order. To the best of our knowledge, this is the first experimental observation in a metallic liquid. Furthermore, previous studies have been performed at much higher temperatures, close to the melting, where the collective motion is diffusive and τ(Q) evolves continuously with $Q^{-2}$ at low Q values.

In order to account for the influence of the shape parameter on the observed relaxation times, Fig. 4e) and f) report the mean relaxation time for both compositions. A slow-down of the dynamics can be observed also in <τ(Q)> for the Pt-based alloy. Surprisingly, the dynamics at the pre-peak, $Q_{PP}$, is even slower than that at low Q values and at the FSDP, indicating the pronounced temporal durability of the medium range structural features in this composition. Previous structural studies suggested that the scattering contribution of the pre-peak arises from Pt-Pt and Pt-Cu partial structure factors [18]. The presence of the pre-peak only in the Pt-based alloy shows that such structural feature is of chemical and thus electronic nature, rather than from a geometrical packing origin, as Pt and Pd can be assumed topologically equal but differ in their electronic structure. Thus, in this scenario, the slow-down of <τ(Q)> at $Q_{pp}$ in the Pt-based alloy would not be related to a preferential fast diffusion path in a slow relaxing matrix as for silicate glasses or other non-metallic glass formers [58,61,62], but would originate from the presence of specific structural motifs that stems from chemical effects. The mechanism is similar to that of de Gennes narrowing at the FSDP usually observed in the frequency domain in high temperature liquids [50], however, here the origin would not be due to geometrical constrains, but rather to strong chemical interactions which reduce the mobility at the mesoscopic scale.

A similar slow-down of the atomic motion in correspondence of a structural pre-peak has been reported for high temperature binary alloys [63]. In such two component systems, a full set of partial structure factors can be experimentally determined, and the presence of a pre-peak has been attributed to the preferential formation of non-equal pairs (as for instance Pt-Cu in our sample), which lead to an isolation of equal atoms and result in a spatial arrangement that can be described as a type of superstructure at the MRO length scale [64]. Although the same structural analysis cannot be done for the quaternary alloys studied here, the structural motifs and the role of the metal-metalloid interactions has been described by simulation studies [65]. The presence of different structural motifs in the two alloys [18], in particular of trigonal prisms in the Pt-based system, somehow leads to different connections between the atomic clusters and to the formation of some favored and temporally stable configurations on the length scales larger than the FSDP in the Pt-alloy.

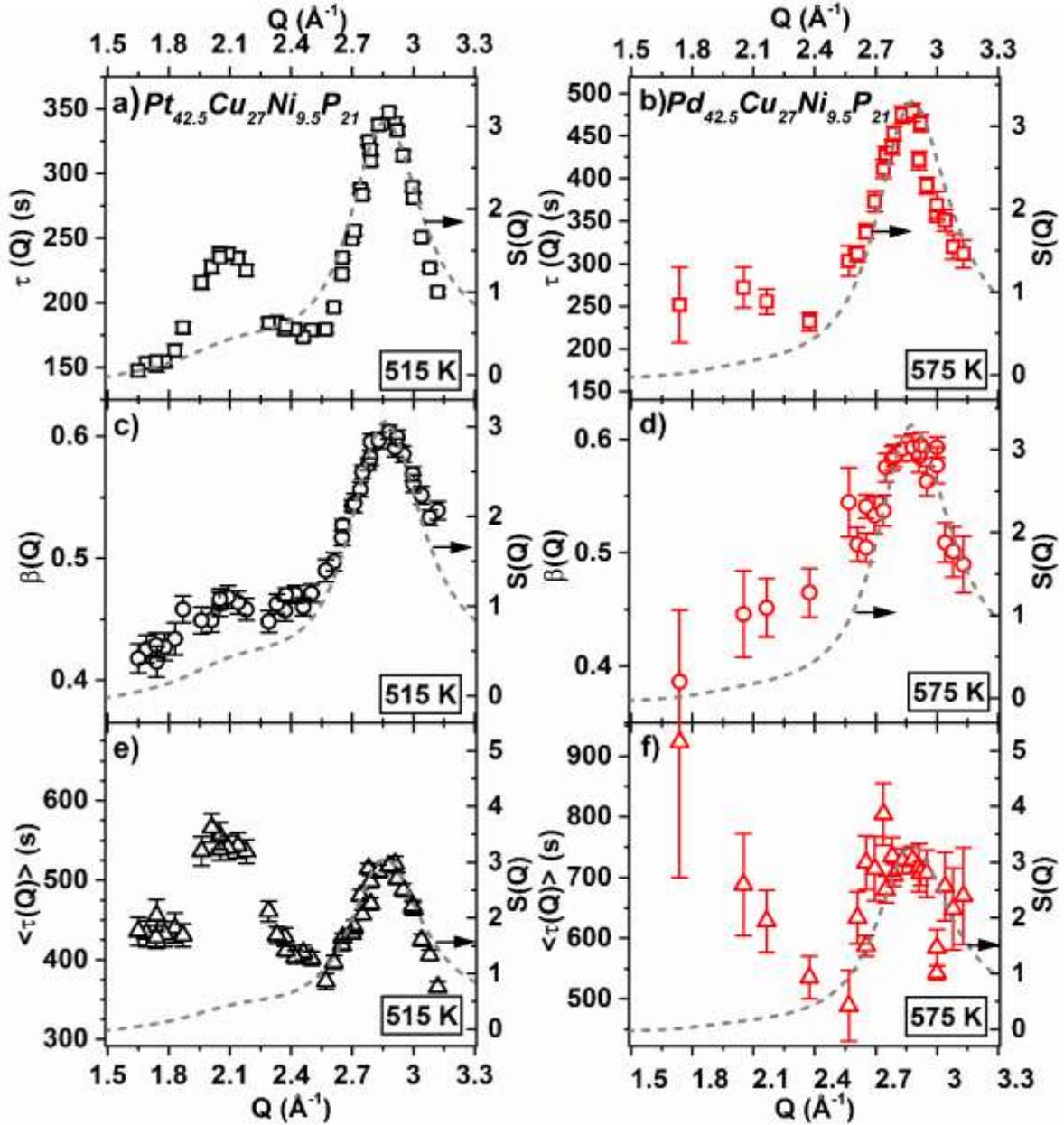

**Figure 4**: **Wave-vector dependence of the dynamics.** Q- dependence of the KWW relaxation time (**a**) and (**b**), KWW exponent (**c**) and (**d**), and average relaxation time (**e**) and (**f**) for the Pt-alloy at T = 515 K (left column, black) and the Pd-alloy at T = 575 K (right column, red). In all panels, the dashed grey lines represent the S(Q) measured with synchrotron XRD.

## Conclusion

In conclusion, our work scrutinizes the microscopic details controlling the evolution of the α-relaxation in supercooled metallic liquids in the vicinity of the glass transition. By comparing the evolution of the α-relaxation process and the structure of two similar supercooled liquids, we identify the different features describing the liquid dynamics. Although the two selected compositions would appear as identical in topological structural models [16], they present different structural motifs on the MRO and SRO length scales [18], which influence their relaxation dynamics. The kinetic fragility - described by the temperature evolution of the microscopic relaxation time - is insensitive to the composition and is similar for both alloys, in agreement with macroscopic studies [44,66]. In contrast, the two liquids have different structural fragilities which mirror the different temperature dependence of the β($Q_{FSDP}$,T) parameter describing the heterogeneous nature of the dynamics. This means that the temperature evolution of dynamical heterogeneities in the supercooled liquid phase reflects the different tendencies of the two liquids to temperature induced structural rearrangements at the MRO length scale. In the case of the $Pd_{42.5}Cu_{27}Ni_{9.5}P_{21}$ alloy, this structure-dynamic relationship results in a relatively weak temperature dependence of the structure and a constant value of β($Q_{FSDP}$,T) in the probed temperature range. Differently, in the Pt-based alloy, the presence of structural rearrangements leads to a continuous evolution of the dynamical heterogeneities during the cooling, as signaled by the increased non-exponential shape of the decay of the density fluctuations on approaching $T_g$. This marked temperature dependence of β($Q_{FSDP}$,T) in the Pt-based melt leads to the failure of the time-temperature superposition principle for the structural relaxation process (Fig. S4 in SI), which contrasts with the usual law of invariance observed in previous studies on metallic glass formers at both microscopic [31–33] and macroscopic [34–36,67] scales. Interestingly, despite the different evolution of the dynamics during cooling, both alloys exhibit similar values of β(Q,T) at the glass transition suggesting the existence of a common degree of heterogeneities in metallic liquids at the dynamical arrest.

In addition, we also find that the wave-vector dependence of the collective motion is influenced by the structure. In the $Pt_{42.5}Cu_{27}Ni_{9.5}P_{21}$ liquid, both τ(Q,T) and β(Q,T) exhibit an unusual evolution with Q with two maxima, one at the position of the FSDP, signature of the well-known increased stability at the interparticle distance, and an anomalous second one at the mesoscopic scale in correspondence with a weak pre-peak in the total structure factor. In contrast, only one maximum at $Q_{FSDP}$ is observed in the Pd-alloy. This different dependence of the dynamics from the probed wave-vector is likely due to a more pronounced MRO in the Pt- than in the Pd-liquid, similarly to what was also reported in high temperature binary alloys [63]. The different MRO in the two systems is likely related to stronger chemical and electronic interactions in the Pt-based alloy than in the Pd-liquid, which also result in the presence of different structural motifs at the level of the SRO. All together our results shed new light on the influence of the structure on the particle motion and the different role played by kinetic and structural fragilities during the vitrification process.

## Materials and Methods

### Materials

The master alloys of $Pt_{42.5}Cu_{27}Ni_{9.5}P_{21}$ and $Pd_{42.5}Cu_{27}Ni_{9.5}P_{21}$ were produced by melting the pure metallic components (purity > 99.95 %) in an arc melter furnace under a Ti-gettered Ar-atmosphere (purity > 99.999%). Afterwards the elemental red P was alloyed inductively with the metallic components in a fused-silica tube under Ar-atmosphere. The alloy then underwent a fluxing treatment in dehydrated pure $B_2O_3$ for at least 6 at 1473 K. This process was used to further purify the melt from possible oxides and impurities and enhances the glass-forming ability of the liquid even further [68]. Ribbons were prepared via melt-spinning by inductive melting of the master-alloy in a fused silica tube and injecting it onto a rotating copper wheel under high-purity Ar atmosphere in a custom-built melt-spinning device.

### X-ray photon correlation spectroscopy (XPCS)

To probe the microscopic dynamics, X-ray photon correlation spectroscopy was performed at the Coherence Applications Beamline P10 at PETRA III at the *Deutsches Elektronen-Synchrotron* (DESY) in Hamburg during two different beamtimes of one week each. A partially coherent beam of 3 x 2 μm² (horizontal x vertical direction) at a photon energy of 8.2 keV and a photon flux on the sample of around $4 \times 10^{10}$ photons/s was used on ribbons with a thickness of ~ 20 μm, mounted in a furnace under vacuum. Speckle patterns were collected with an EIGER X4M detector 1.8 m downstream of the sample at the end of a horizontally rotatable diffractometer arm, enabling measurements at different wavevectors Q. The low Q limit was set at ~ 1.6 Å$^{-1}$ due to the weak scattering intensity at small angles, while the maximum angle that was reachable with the experimental setup was at 43°, corresponding to Q = 3.06 Å$^{-1}$, limiting thus the resolution of the FSDP on the high-Q flank. The wave-vector dependence of the dynamics was measured at different temperatures in the supercooled liquid phase. The exposure time per frame was adjusted to the observed timescales ranging from 0.1 s to 0.5 s. The overall measurement time was also adjusted at each temperature and wavevector position assuring the observation of a complete decorrelation of the signal, up to a maximum of ~ $10^4$ s for the lowest measured temperature. The detector covers an angle of approximately 3° which corresponds to a ΔQ of about ~ 0.2 Å$^{-1}$. By binning the area of the detector, we could therefore probe up to 8 Qs simultaneously for each detector position.

Data were analyzed following the procedure reported in Ref. [69]. For each Q and temperature, we calculated the two times correlation function (TTCF) at two different times, $t_1$ and $t_2$:

$$G(Q, t_1, t_2) = \frac{<I(Q,t_1) \cdot I(Q,t_2)>_p}{<I(Q,t_1)>_p \cdot <I(Q,t_2)>_p}, \quad (1)$$

where $<...>_p$ represents the average over all detector pixels corresponding to the same Q-range. Intensity auto correlation functions $g_2(Q,t)$ were calculated by averaging the TTCF over the whole measured temporal interval. The $g_2(Q,t)$ functions are directly related to the intermediate scattering function, $f(Q,t)$, through the Siegert relation $g_2(Q,t) = 1 + \gamma \cdot |f(Q,t)|^2$ with $\gamma$ being the experimental contrast. The $g_2(Q,t)$ functions were modelled using a Kohlrausch-Williams-Watts (KWW) function:

$$g_2(Q,t,T) = 1 + c(Q,T) \exp\left(-2\left(\frac{t}{\tau(Q,T)}\right)^{\beta(Q,T)}\right), \quad (2)$$

where $\tau(Q,T)$ is the structural relaxation time, $\beta(Q,T)$ is the shape parameter and $c(Q,T)$ is the product of the experimental contrast $\gamma(Q,T)$ and the square of the non-ergodicity factor $f_q(Q,T)$. The samples were first equilibrated at about $T_g+30$ K to remove their thermal history, and then slowly cooled with 5 K min$^{-1}$ to the measurement temperatures. At each temperature, we waited about 10 min for temperature equilibration before starting the data acquisition. The amorphous structure was checked during the experiment and no signs of crystallization were observed, in agreement with the time-temperature-transformation diagram in Ref. [22,70]. The temperature calibration of the furnace was checked by crystallizing additional samples and comparing the crystallization temperature $T_x$ of the $Pt_{42.5}Cu_{27}Ni_{9.5}P_{21}$ and $Pd_{42.5}Cu_{27}Ni_{9.5}P_{21}$ samples at a constant heating rate of 5 K min$^{-1}$ obtained in differential scanning calorimetry (DSC).

For the solute-rich metallic liquid $Pt_{42.5}Cu_{27}Ni_{9.5}P_{21}$, the main contributions to the pre-peak originate from the Pt-Pt and Pt-Cu partial structure factors, which, due to their high form factor, dominate the overall structure factor S(Q), while the signal of the Pd-based alloy is dominated by the Pd-Pd and Pd-Cu correlations (see Supplementary Information of Ref. [18] for further details).

### High energy synchrotron X-ray scattering (HE-XRD)

In-situ X-ray scattering experiments have been performed at the beamline P02.1 at PETRA III at the Deutsches Elektronen Synchrotron (DESY) in Hamburg [71]. For the measurements in transmission mode a wavelength of 0.207 Å (60 keV) and a beam size of 0.8 × 0.8 mm² was used. The samples were attached on a solid Ag-block of a THMS-600 LINKAM furnace using a Cu-paste and heated at a rate of 0.33 Ks$^{-1}$ under a constant flow of high purity Ar (Ar 6.0).

Prior to the measurements the samples were heated at 0.33 K s$^{-1}$ to the supercooled liquid state ($T_{g,end}$ + 10 K) and subsequently cooled from this point at the same rate of 0.33 Ks$^{-1}$. For the acquisition of the intensity patterns a Perkin Elmer XRD1621 CsI bonded amorphous silicon detector (2048 pixels × 2048 pixels) was used. The integration of the dark-subtracted, two-dimensional X-ray diffraction patterns was performed with the Fit2D data analysis software [72]. Further processing of the intensity data was realized with the PDFgetX2 software [73]. The background scattering was assumed to be constant with temperature and mainly originating from the setup. It was measured at room temperature and was subtracted from the integrated intensity data. For the corrections of the raw data, sample absorption, polarization and multiple scattering were considered. The total structure factor S(Q) was calculated as [74]

$$S(Q) = 1 + \frac{I_C(Q) - \langle f(Q)^2 \rangle}{\langle f(Q) \rangle^2}, \tag{3}$$

where $I_C(Q)$ is the coherently scattered intensity and f(Q) is the atomic form factor. The angle brackets denote a compositional average over all constituents. S(Q) contains all the structural information and is composed of n(n+1)/2 partial structure factors [75],

$$S(Q) = \sum_{i \leq j} w_{ij} S_{ij}(Q), \tag{4}$$

where $w_{ij}$ is the weighting factor expressed as

$$w_{ij} = \frac{c_i c_j f_i(Q) f_j(Q)}{\langle f(Q) \rangle^2} \tag{5}$$

where $c_i$ and $c_j$ are the molar concentration of element i and j.

The Fourier transform of the total structure factor yields the reduced pair distribution function,

$$G(r) = \frac{2}{\pi} \int_0^\infty Q[S(Q) - 1] \sin(Qr) \, dQ, \tag{6}$$

where r is the distance to the reference atom. Each G(r) pattern was optimized using an optimization algorithm in PDFgetX2 as described in Ref. [76]. An upper limit of Q = 14.5 Å$^{-1}$ was used here for the transformation. This $Q_{max}$ ensures a successful evaluation of the data in real-space without the loss of significant structural details, as already observed by Ma et al. for even lower $Q_{max}$-values [47].

**Thermomechanical analysis**

To determine the kinetic fragility on a macroscopic experiment three-point beam bending (3PBB) was performed using a NETZSCH TMA 402 F3 thermomechanical analyzer under a static loading force of 10 N. The beams were cut out of fully X-ray-amorphous plates with the dimensions of 1.5 × 13 × 40 mm. For a good signal-to-noise ratio the thickness of the beams was varied from 0.4 to 1.5 mm. The viscosity η can be derived from the deflection rate ů(t), applying the equation [77]

$$\eta(t) = -\frac{g \cdot L^3}{144\, I\dot{u}} \cdot \left[M + \frac{\rho A L}{1.6}\right], \quad (7)$$

with the applied load M and density ρ combined with the geometric information of the sample (cross-sectional area of the beam A, cross-sectional moment of inertia I and distance between the supporting edges of the machine L).

Isothermal measurements between 533 K and 573 K, as well as measurements with a constant heating rate of $q_h = 0.333$ Ks$^{-1}$ were carried out. During the isothermal measurements the same heating rate of 0.333 Ks$^{-1}$ was applied to reach the desired isothermal plateau temperature. A more detailed description of the technique and calculations to obtain the viscosity from the deflection rate can be found in Refs. [19,66].

The evolution of the equilibrium viscosities over temperature can be described with the empirical Vogel-Fulcher-Tammann (VFT) equation [11]

$$\eta(T) = \eta_0 \cdot \exp\left(\frac{D^* \cdot T_0}{T - T_0}\right), \quad (8)$$

with D$^*$ being the fragility parameter and T$_0$ being the VFT temperature, where viscosity would be diverging. The parameter η$_0$ corresponds to the lowest possible viscosity in the high temperature liquid state.

# References

1. Donth, E. *The glass transition: relaxation dynamics in liquids and disordered materials*. vol. 48 (Springer Science & Business Media, 2013).

2. Anderson, P. W. Through the Glass Lightly. *Science (80-. ).* **267**, 1615–1616 (1995).

3. Berthier, L. & Biroli, G. Theoretical perspective on the glass transition and amorphous materials. *Rev. Mod. Phys.* **83**, 587–645 (2011).

4. Ediger, M. D. & Harrowell, P. Perspective: Supercooled liquids and glasses. *J. Chem. Phys.* **137**, (2012).

5. Dyre, J. C. Colloquium: The glass transition and elastic models of glass-forming liquids. *Rev. Mod. Phys.* **78**, 953–972 (2006).

6. Chandler, D. & Garrahan, J. P. Dynamics on the way to forming glass: Bubbles in space-time. *Annu. Rev. Phys. Chem.* **61**, 191–217 (2010).

7. Tong, H. & Tanaka, H. Structural order as a genuine control parameter of dynamics in simple glass formers. *Nat. Commun.* **10**, 4–6 (2019).

8. Tanaka, H., Kawasaki, T., Shintani, H. & Watanabe, K. Critical-like behaviour of glass-forming liquids. *Nat. Mater.* **9**, 324–331 (2010).

9. Böhmer, R. *et al.* Nonexponential relaxations in strong and fragile glass formers. *J. Chem. Phys.* **99**, 4201–4209 (1993).

10. Ediger, M. D. Spatially Heterogeneous Dynamics in Supercooled Liquids. *Annu. Rev. Phys. Chem.* **51**, 99–128 (2000).

11. Angell, C. A. Formation of glasses from liquids and biopolymers. *Science (80-. ).* **267**, 1924–1935 (1995).

12. Scopigno, T., Ruocco, G., Sette, F. & Monaco, G. Is the Fragility of a Liquid Embedded in the Properties of Its Glass? *Science (80-. ).* **302**, 849–852 (2003).

13. Angell, C. A. & Martinez, L.-M. A thermodynamic connection to the fragility of glass-forming liquids. *Nature* **410**, 663–667 (2001).

14. Novikov, V. N. & Sokolov, A. P. Poisson's ratio and the fragility of glass-forming

liquids. *Nature* **431**, 961–963 (2004).

15. Wei, S. *et al.* Linking structure to fragility in bulk metallic glass-forming liquids. *Appl. Phys. Lett.* **106**, 10–15 (2015).

16. Miracle, D. B. A structural model for metallic glasses. *Microsc. Microanal.* **10**, 786–787 (2004).

17. Miracle, D. B. The efficient cluster packing model - An atomic structural model for metallic glasses. *Acta Mater.* **54**, 4317–4336 (2006).

18. Gross, O. *et al.* Signatures of structural differences in Pt–P- and Pd–P-based bulk glass-forming liquids. *Commun. Phys.* **2**, 83 (2019).

19. Gross, O. *et al.* The kinetic fragility of Pt-P- and Ni-P-based bulk glass-forming liquids and its thermodynamic and structural signature. *Acta Mater.* **132**, 118–127 (2017).

20. Kato, H. *et al.* Fragility and thermal stability of Pt- and Pd-based bulk glass forming liquids and their correlation with deformability. *Scr. Mater.* **54**, 2023–2027 (2006).

21. Gallino, I., Schroers, J. & Busch, R. Kinetic and thermodynamic studies of the fragility of bulk metallic glass forming liquids. *J. Appl. Phys.* **108**, 063501 (2010).

22. Gross, O. *et al.* On the high glass-forming ability of Pt-Cu-Ni/Co-P-based liquids. *Acta Mater.* **141**, 109–119 (2017).

23. Neuber, N. *et al.* On the thermodynamics and its connection to structure in the Pt-Pd-Cu-Ni-P bulk metallic glass forming system. *Acta Mater.* **220**, 117300 (2021).

24. Schroers, J. & Johnson, W. L. Ductile bulk metallic glass. *Phys. Rev. Lett.* **93**, 255506 (2004).

25. Kumar, G., Neibecker, P., Liu, Y. H. & Schroers, J. Critical fictive temperature for plasticity in metallic glasses. *Nat. Commun.* **4**, 1536 (2013).

26. Kumar, G., Prades-Rodel, S., Blatter, A. & Schroers, J. Unusual brittle behavior of Pd-based bulk metallic glass. *Scr. Mater.* **65**, 585–587 (2011).

27. Schroers, J. & Johnson, W. L. Highly processable bulk metallic glass-forming alloys in the Pt-Co-Ni-Cu-P system. *Appl. Phys. Lett.* **84**, 3666–3668 (2004).

28. Nishiyama, N. *et al.* The world's biggest glassy alloy ever made. *Intermetallics* **30**, 19–24 (2012).

29. Madsen, A., Fluerasu, A. & Ruta, B. Structural dynamics of materials probed by X-ray photon correlation spectroscopy. *Synchrotron Light Sources Free. Lasers Accel. Physics, Instrum. Sci. Appl.* 1989–2018 (2020) doi:10.1007/978-3-030-23201-6_29.

30. Boon, J. P. & Yip, S. *Molecular hydrodynamics*. (Courier Corporation, 1991).

31. Amini, N. *et al.* Intrinsic relaxation in a supercooled ZrTiNiCuBe glass forming liquid. *Phys. Rev. Mater.* **5**, 1–8 (2021).

32. Ruta, B. *et al.* Atomic-scale relaxation dynamics and aging in a metallic glass probed by X-ray photon correlation spectroscopy. *Phys. Rev. Lett.* **109**, 1–5 (2012).

33. Hechler, S. *et al.* Microscopic evidence of the connection between liquid-liquid transition and dynamical crossover in an ultraviscous metallic glass former. *Phys. Rev. Mater.* **2**, 1–6 (2018).

34. Wang, L. M., Liu, R. & Wang, W. H. Relaxation time dispersions in glass forming metallic liquids and glasses. *J. Chem. Phys.* **128**, (2008).

35. Qiao, J. C. & Pelletier, J. M. Dynamic universal characteristic of the main (α) relaxation in bulk metallic glasses. *J. Alloys Compd.* **589**, 263–270 (2014).

36. Yao, Z. F., Qiao, J. C., Pelletier, J. M. & Yao, Y. Characterization and modeling of dynamic relaxation of a Zr-based bulk metallic glass. *J. Alloys Compd.* **690**, 212–220 (2017).

37. Wang, W. H. Dynamic relaxations and relaxation-property relationships in metallic glasses. *Prog. Mater. Sci.* **106**, 100561 (2019).

38. Qiao, J. C. *et al.* Structural heterogeneities and mechanical behavior of amorphous alloys. *Prog. Mater. Sci.* **104**, 250–329 (2019).

39. Zhang, L. T. *et al.* Dynamic mechanical relaxation and thermal creep of high-entropy La30Ce30Ni10Al20Co10 bulk metallic glass. *Sci. China Physics, Mech. Astron.* **64**, (2021).

40. Luo, P., Wen, P., Bai, H. Y., Ruta, B. & Wang, W. H. Relaxation Decoupling in

Metallic Glasses at Low Temperatures. *Phys. Rev. Lett.* **118**, 1–6 (2017).

41. Wang, Z., Sun, B. A., Bai, H. Y. & Wang, W. H. Evolution of hidden localized flow during glass-to-liquid transition in metallic glass. *Nat. Commun.* **5**, (2014).

42. Sokolov, A. P., Kisliuk, A., Soltwisch, M. & Quitmann, D. Medium-range order in glasses: Comparison of Raman and diffraction measurements. *Phys. Rev. Lett.* **69**, 1540–1543 (1992).

43. Meyer, A., Busch, R. & Schober, H. Time-temperature superposition of structural relaxation in a viscous metallic liquid. *Phys. Rev. Lett.* **83**, 5027–5029 (1999).

44. Frey, M. *et al.* Determining the fragility of bulk metallic glass forming liquids via modulated DSC. *J. Phys. Condens. Matter* **32**, 324004 (2020).

45. Gross, O. Precious metal based bulk glass-forming liquids: Development, thermodynamics, kinetics and structure. *Diss. Saarl. Univ.* (2018) doi:10.22028/D291-27993.

46. Fabian, R. & Sidebottom, D. L. Dynamic light scattering in network-forming sodium ultraphosphate liquids near the glass transition. *Phys. Rev. B - Condens. Matter Mater. Phys.* **80**, 1–7 (2009).

47. Ma, D., Stoica, A. D. & Wang, X. L. Power-law scaling and fractal nature of medium-range order in metallic glasses. *Nat. Mater.* **8**, 30–34 (2009).

48. Gallino, I. On the fragility of bulk metallic glass forming liquids. *Entropy* **19**, (2017).

49. Adam, G. & Gibbs, J. H. On the temperature dependence of cooperative relaxation properties in glass-forming liquids. *J. Chem. Phys.* **43**, 139–146 (1965).

50. De Gennes, P. G. Liquid dynamics and inelastic scattering of neutrons. *Physica* **25**, 825–839 (1959).

51. Yang, F., Kordel, T., Holland-Moritz, D., Unruh, T. & Meyer, A. Structural relaxation as seen by quasielastic neutron scattering on viscous Zr-Ti-Cu-Ni-Be droplets. *J. Phys. Condens. Matter* **23**, (2011).

52. Fuchs, M., Hofacker, I. & Latz, A. Primary relaxation in a hard-sphere system. *Phys. Rev. A* **45**, 898–912 (1992).

53. Sciortino, F., Fabbian, L., Chen, S. H. & Tartaglia, P. Supercooled water and the kinetic glass transition. II. Collective dynamics. *Phys. Rev. E - Stat. Physics, Plasmas, Fluids, Relat. Interdiscip. Top.* **56**, 5397–5404 (1997).

54. Weysser, F., Puertas, A. M., Fuchs, M. & Voigtmann, T. Structural relaxation of polydisperse hard spheres: Comparison of the mode-coupling theory to a Langevin dynamics simulation. *Phys. Rev. E - Stat. Nonlinear, Soft Matter Phys.* **82**, 1–21 (2010).

55. Handle, P. H., Rovigatti, L. & Sciortino, F. Q-Independent Slow Dynamics in Atomic and Molecular Systems. *Phys. Rev. Lett.* **122**, 175501 (2019).

56. Poon, W. Colloids as Big Atoms. *Science (80-. ).* **304**, 830–831 (2004).

57. Colmenero, J. & Arbe, A. Recent progress on polymer dynamics by neutron scattering: From simple polymers to complex materials. *J. Polym. Sci. Part B Polym. Phys.* **51**, 87–113 (2013).

58. Ruta, B. *et al.* Revealing the fast atomic motion of network glasses. *Nat. Commun.* **5**, (2014).

59. Novikov, V. N., Schweizer, K. S. & Sokolov, A. P. Coherent neutron scattering and collective dynamics on mesoscale. *J. Chem. Phys.* **138**, (2013).

60. Tolle Albert. Neutron scattering studies of the model glass former ortho -terphenyl. *Rep. Prog. Phys.* **64**, 1473 (2001).

61. Horbach, J., Kob, W. & Binder, K. Dynamics of Sodium in Sodium Disilicate: Channel Relaxation and Sodium Diffusion. *Phys. Rev. Lett.* **88**, 4 (2002).

62. Meyer, A., Horbach, J., Kob, W., Kargl, F. & Schober, H. Channel formation and intermediate range order in sodium silicate melts and glasses. *Phys. Rev. Lett.* **93**, 1–4 (2004).

63. Voigtmann, T. *et al.* Atomic diffusion mechanisms in a binary metallic melt. *Epl* **82**, 1–6 (2008).

64. Nowak, B. *et al.* Partial structure factors reveal atomic dynamics in metallic alloy melts. *Phys. Rev. Mater.* **1**, 3–7 (2017).

65. Guan, P. F., Fujita, T., Hirata, A., Liu, Y. H. & Chen, M. W. Structural origins of the excellent glass forming ability of Pd 40Ni 40P 20. *Phys. Rev. Lett.* **108**, 1–5 (2012).

66. Neuber, N. *et al.* The role of Ga addition on the thermodynamics, kinetics, and tarnishing properties of the Au-Ag-Pd-Cu-Si bulk metallic glass forming system. *Acta Mater.* **165**, 315–326 (2019).

67. Soriano, D. *et al.* Relaxation dynamics of Pd-Ni-P metallic glass: Decoupling of anelastic and viscous processes. *J. Phys. Condens. Matter* **33**, (2021).

68. Kui, H. W., Greer, A. L. & Turnbull, D. Formation of bulk metallic glass by fluxing. *Appl. Phys. Lett.* **45**, 615–616 (1984).

69. Chushkin, Y., Caronna, C. & Madsen, A. A novel event correlation scheme for X-ray photon correlation spectroscopy. *J. Appl. Crystallogr.* **45**, 807–813 (2012).

70. Schroers, J., Wu, Y., Busch, R. & Johnson, W. L. Transition from nucleation controlled to growth controlled crystallization in Pd43Ni10Cu27P20 melts. *Acta Mater.* **49**, 2773–2781 (2001).

71. Dippel, A. C. *et al.* Beamline P02.1 at PETRA III for high-resolution and high-energy powder diffraction. *J. Synchrotron Radiat.* **22**, 675–687 (2015).

72. Hammersley, A. P. {FIT2D}: An Introduction and Overview. *Eur. Sychrotron Radi Facil. Int. Rep. ESRF97HA02T* **68**, (1997).

73. Qiu, X., Thompson, J. W. & Billinge, S. J. L. PDFgetX2: A GUI-driven program to obtain the pair distribution function from X-ray powder diffraction data. *J. Appl. Crystallogr.* **37**, 678 (2004).

74. Egami, T. & Billinge, S. J. L. *Underneath the Bragg Peaks: structural analysis of complex materials*. *Pergamon Materials Series* vol. 7 (2003).

75. Faber, T. E. & Ziman, J. M. A theory of the electrical properties of liquid metals. *Philos. Mag.* **11**, 153–173 (1965).

76. Wei, S. *et al.* Structural evolution on medium-range-order during the fragile-strong transition in Ge15Te85. *Acta Mater.* **129**, 259–267 (2017).

77. Hagy, H. E. Experimental Evaluation of Beam???Bending Method of Determining

Glass Viscosities in the Range 108 to 1015 Poises. *J. Am. Ceram. Soc.* **46**, 93–97 (1963).

## Acknowledgements

We acknowledge DESY (Hamburg, Germany), a member of the Helmholtz Association HGF, for the provision of experimental facilities. Parts of this research were carried out at PETRA III beamline P02.1 and we would like to thank Alexander Schoekel and Michael Wharmby for assistance. Further we want to thank our colleagues B. Adam, L. Ruschel, S.S. Riegler, and H. Jiang for collaboration and fruitful discussions concerning the topic. This project has received funding from the European Research Council (ERC) under the European Union's Horizon 2020 research and innovation programme (Grant Agreement No 948780). The authors declare that they have no competing interests. All data needed to evaluate the conclusions in the paper are present in the paper and/or the Supplementary Materials.

## Author Contributions
N.N., R.B., O.G. and B.R. conceived the study. N.N. and M.F. prepared the samples. N.N., O.G., B.B., A.K., S.H., I.G., B.R., E. P., F. Y. and M.F. planned and conducted the synchrotron X-ray experiments with the help of F.W. and M.S. N.N. and B.R. analyzed the synchrotron data. N.N. and M.F. conducted and analyzed the thermo-mechanical and calorimetric experiments. N.N. and B.R. wrote the paper with input from R.B., E.P., F.Y., I.G and O.G. All authors proofread the article and contributed extensively to the discussion.

## Supplementary Information

It shall be noted that large portions of the synchrotron radiation-based experiments reported in this work (X-ray photon correlation spectroscopy, as well as total diffraction experiments) were repeated on several beamtimes, reproducing the given results.

## Supplementary Figures

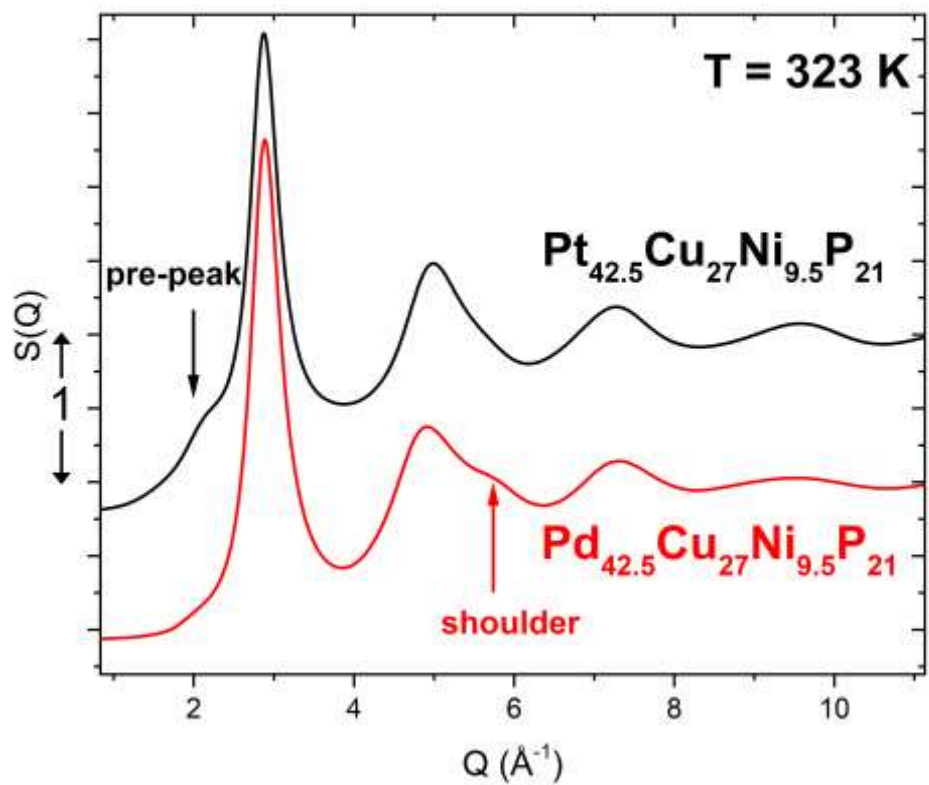

**SI Figure S1:** Static structure factor S(Q) of the glassy state at 323 K of the two probed alloys $Pt_{42.5}Cu_{27}Ni_{9.5}P_{21}$ and $Pd_{42.5}Cu_{27}Ni_{9.5}P_{21}$ representing their structural differences. The S(Q) of the $Pt_{42.5}Cu_{27}Ni_{9.5}P_{21}$ system features a significant pre-peak at low Q around 2 Å$^{-1}$, indicating a significant medium-range order, whereas the static structure factor of $Pd_{42.5}Cu_{27}Ni_{9.5}P_{21}$ has a shoulder at the 2$^{nd}$ diffraction peak between 5 and 6 Å$^{-1}$, which is suggesting an icosahedral short-range order.

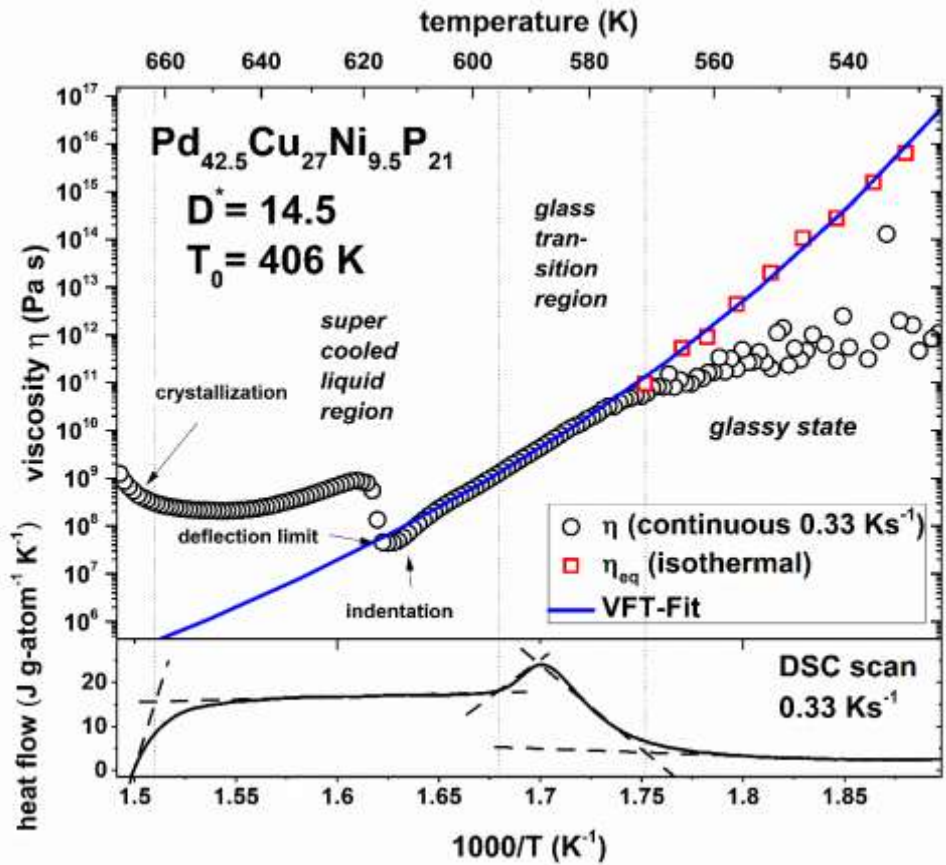

**SI Figure S2:** Equilibrium viscosity of $Pd_{42.5}Cu_{27}Ni_{9.5}P_{21}$ as a function of inverse temperature determined by three-point beam bending, following the methodology described in detail in Ref. [66]. The black open circles mark the results from a continuous measurement at 0.33 K s$^{-1}$, while the red open squares mark the equilibrium viscosities obtained in isothermal measurements. The data corresponding to the equilibrium viscosities is used for the Vogel-Fulcher-Tammann (VTF)-fit resulting in a fragility parameter $D^*$ of 14.5 and a $T_0$ of 406 K. For a better understanding of the respective temperature region of the different states of glass, supercooled liquid and crystallization a thermogram obtained by differential scanning calorimetry (DSC) at 0.33 K s$^{-1}$ is added below.

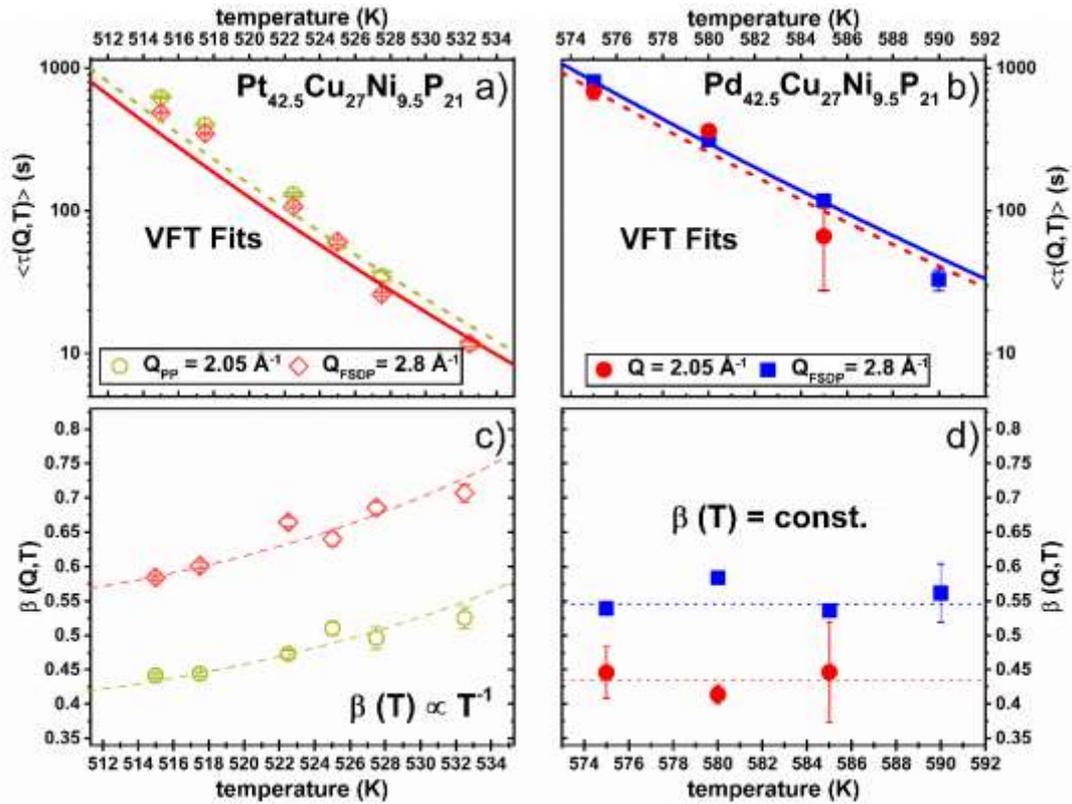

**SI Figure S3:** **(a)** and **(b)** $\langle\tau(Q,T)\rangle$ of the Pt- and Pd-based alloys respectively measured at $Q_{FSDP}$ and $Q = 2.05$ Å$^{-1}$ corresponding to the position of the maximum of the pre-peak in the S(Q) of the Pt-alloy. The solid and dashed lines are fits based on the VFT equation, using the macroscopic fragility parameter D* and the VFT temperature $T_0$, (Ref. [19] for Pt and SI Fig. S2 for Pd) while leaving the high temperature relaxation time $\tau_0$ a fitting parameter for each set of data (SI Table 1). **(c)** and **(d)** Temperature dependence of the corresponding $\beta(Q,T)$ parameters of $Pt_{42.5}Cu_{27}Ni_{9.5}P_{21}$ **(c)** and $Pd_{42.5}Cu_{27}Ni_{9.5}P_{21}$ **(d)**. The dashed and dotted lines are guides to the eyes.

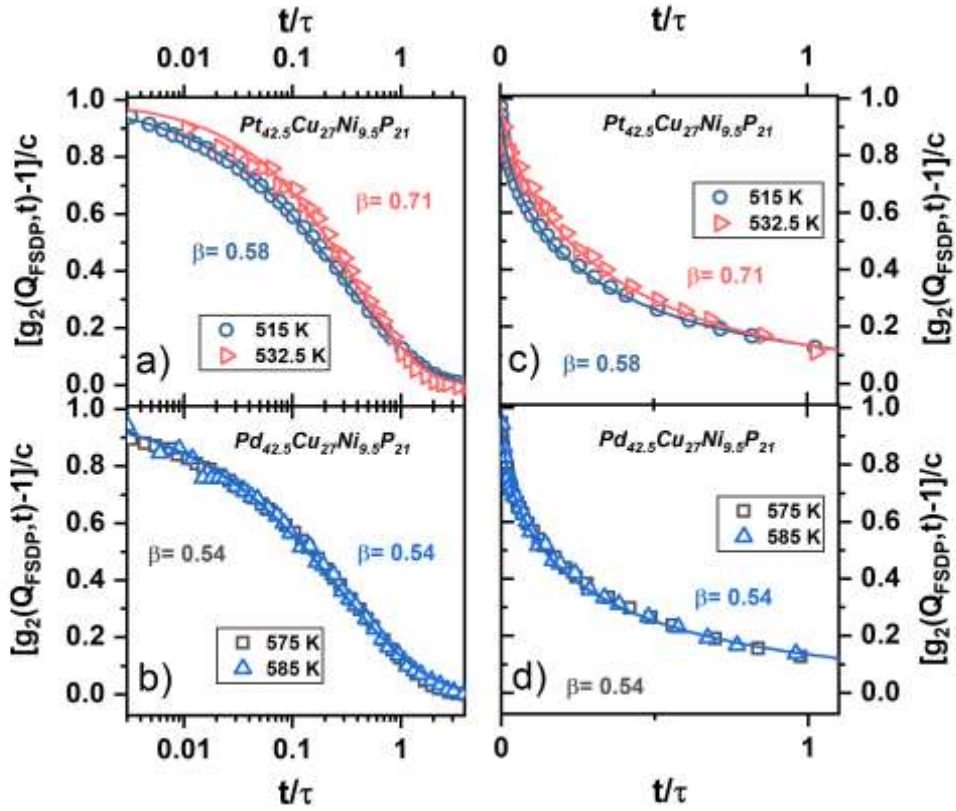

**SI Figure S4:** Temperature dependence of normalized intensity autocorrelation functions measured with XPCS for $Pt_{42.5}Cu_{27}Ni_{9.5}P_{21}$ **(a)**, and $Pd_{42.5}Cu_{27}Ni_{9.5}P_{21}$ **(b)** at the position of the structural FSDP ($Q_{FSDP} \approx 2.8$ Å$^{-1}$ for both systems). Two representative correlation curves at two different temperatures are reported as a function of $t/\tau$ for $Pt_{42.5}Cu_{27}Ni_{9.5}P_{21}$ **(a)** and $Pd_{42.5}Cu_{27}Ni_{9.5}P_{21}$ **(b)**, showing the failure **(a)** and the validity **(b)** of the time-temperature superposition principle. For reasons of clarity the same data of **(a)** and **(b)** is shown also on a linear scale of $t/\tau$ in panels **(c)** and **(d),** respectively.

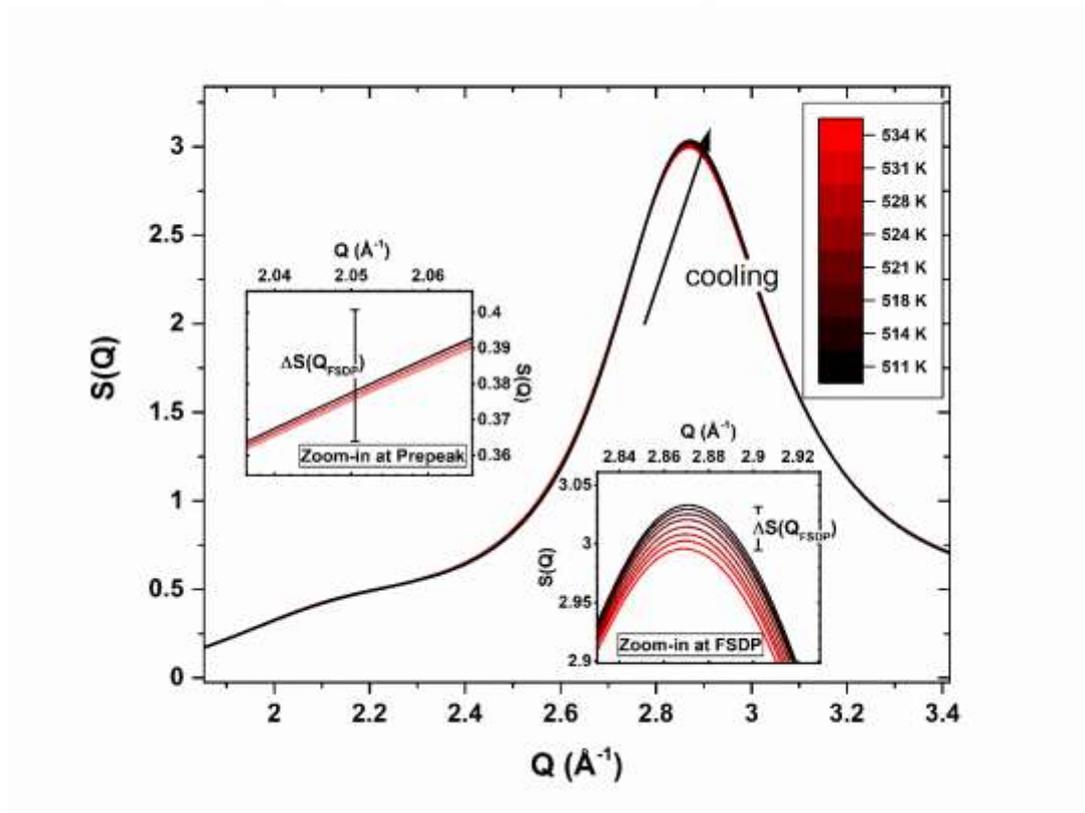

**SI Figure S5:** Evolution of the total static structure factor S(Q) of the Pt-based alloy in the same temperature interval as observed in XPCS (~ 532.5 K-510.5 K), measured while cooling from the supercooled liquid at a rate of 0.33 K s$^{-1}$. Two of the insets show a zoom-in at the FSDP and the prepeak, unveiling similar changes in the S(Q) at the FSDP and the prepeak positions.

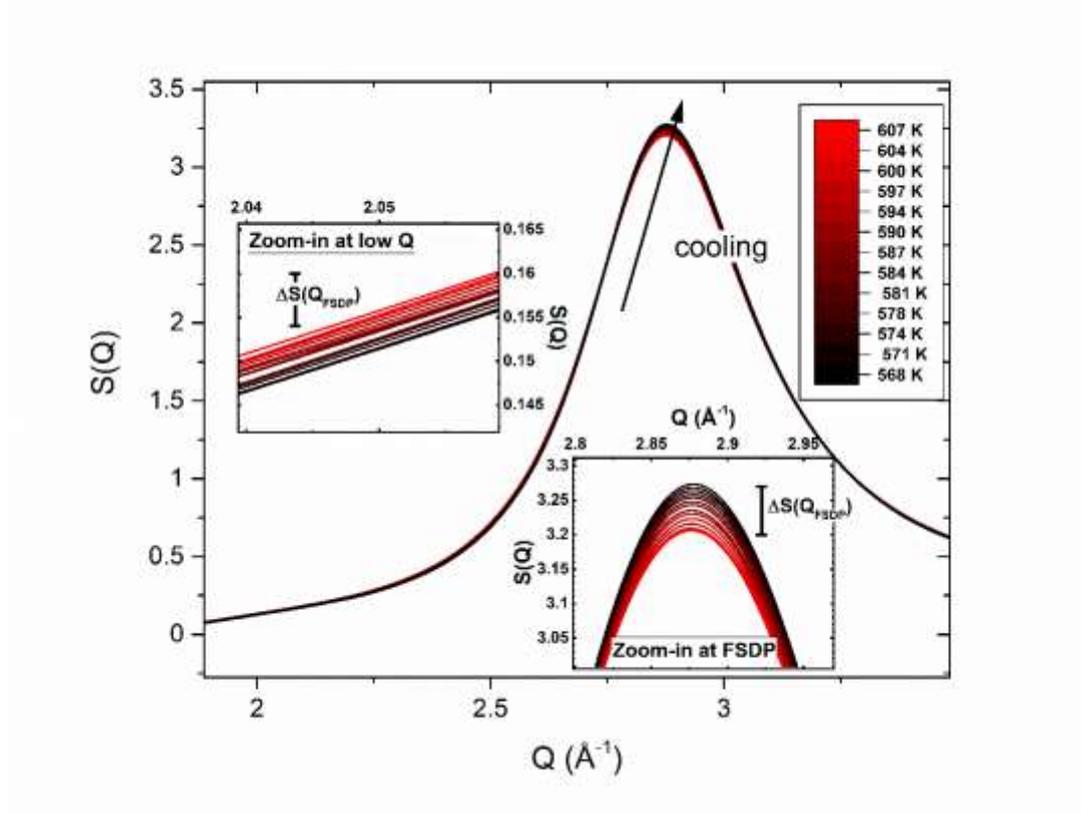

**SI Figure S6:** Evolution of the total static structure factor S(Q) of the Pd-based alloy in the same temperature interval as observed in XPCS ( ~ 600 K-570 K), measured while cooling from the supercooled liquid at a rate of 0.33 K s$^{-1}$. The two insets show a zoom-in at the FSDP and the low Q region

**SI Table 1:** Fitting parameters of relaxation times obtained in XPCS when using fixed D$^*$ and T$_0$ VFT-parameters

|  | *Pt$_{42.5}$Cu$_{27}$Ni$_{9.5}$P$_{21}$* | *Pd$_{42.5}$Cu$_{27}$Ni$_{9.5}$P$_{21}$* |
| --- | --- | --- |
| ***Q = 2.05 Å$^{-1}$*** | $\tau_0 = 10 \cdot 10^{-13} \pm 1.3 \cdot 10^{-13}$ s | $\tau_0 = 5.2 \cdot 10^{-13} \pm 0.34 \cdot 10^{-13}$ s |
| ***Q$_{FSDP}$ = 2.8 Å$^{-1}$*** | $\tau_0 = 8.4 \cdot 10^{-13} \pm 0.7 \cdot 10^{-13}$ s | $\tau_0 = 6 \cdot 10^{-13} \pm 0.05 \cdot 10^{-13}$ s |
| ***VFT Parameters*** | D$^*$ = 15.3 , T$_0$ = 354 K | D$^*$ = 14.5 , T$_0$ = 406 K |